\documentclass[trackchanges]{aastex7}

\newcommand{\civ}{C\,\textsc{iv}}
\hypersetup{linkcolor=red,citecolor=cyan,filecolor=blue,urlcolor=magenta}
\usepackage{subfigure}
\usepackage{xcolor}
\usepackage{verbatim}
\shorttitle{B3 1239+376 at $z$ = 3.82 as a $\gamma$-ray emitter.}
\shortauthors{Zhang et al.}
\begin{document}

\title{The $\gamma$-ray-emitting blazar B3 1239+376 at $z$ = 3.82 identified in a multi-wavelength context}

\correspondingauthor{Neng-Hui Liao}
\email{nhliao@gzu.edu.cn}

\author{Wei Zhang}
\affiliation{College of Physics, Guizhou University, Guiyang 550025, People’s Republic of China}
\email{xxx}  

\author[0000-0001-6614-3344]{Neng-Hui Liao} 
\affiliation{College of Physics, Guizhou University, Guiyang 550025, People’s Republic of China}
\email{nhliao@gzu.edu.cn}

\author[]{Hao-Yi Huang} 
\affiliation{College of Physics, Guizhou University, Guiyang 550025, People’s Republic of China}
\email{xxx}

\author[]{Hai Lei} 
\affiliation{School of Physics and Materials Science, Guangzhou University, Guangzhou 510006, People’s Republic of China}
\email{xxx}

\author[]{Xiong Jiang} 
\affiliation{Key Laboratory of Dark Matter and Space Astronomy, Purple Mountain Observatory, Chinese Academy of Sciences, Nanjing 210023, People’s Republic of China}
\affiliation{School of Astronomy and Space Science, University of Science and Technology of China, Hefei, Anhui 230026, People’s Republic of China}
\email{xxx}

\author[0000-0002-7152-3621]{Ning Jiang}
\affiliation{School of Astronomy and Space Science, University of Science and Technology of China, Hefei, Anhui 230026, People’s Republic of China}
\affiliation{Key Laboratory for Research in Galaxies and Cosmology, Department of Astronomy, The University of Science and Technology of China, Chinese Academy of Sciences, Hefei, Anhui 230026, People’s Republic of China}
\email{xxx} 

\author{Zhen-Feng Sheng}
\affiliation{Institute of Deep Space Sciences, Deep Space Exploration Laboratory, Hefei 230026, People’s Republic of China}
\email{xxx}

\author{Lu-Lu Fan}
\affiliation{College of Physics, Guizhou University, Guiyang 550025, People’s Republic of China}
\affiliation{School of Astronomy and Space Science, University of Science and Technology of China, Hefei, Anhui 230026, People’s Republic of China}
\affiliation{Key Laboratory for Research in Galaxies and Cosmology, Department of Astronomy, The University of Science and Technology of China, Chinese Academy of Sciences, Hefei, Anhui 230026, People’s Republic of China}
\affiliation{Institute of Deep Space Sciences, Deep Space Exploration Laboratory, Hefei 230026, People’s Republic of China}
\email{xxx}

\author[0000-0002-1517-6792]{Ting-Gui Wang}
\affiliation{School of Astronomy and Space Science, University of Science and Technology of China, Hefei, Anhui 230026, People’s Republic of China}
\affiliation{Key Laboratory for Research in Galaxies and Cosmology, Department of Astronomy, The University of Science and Technology of China, Chinese Academy of Sciences, Hefei, Anhui 230026, People’s Republic of China}
\email{xxx} 
%% Use the \collaboration command to identify collaborations. This command
%% takes an optional argument that is either a number or the word "all"
%% which tells the compiler how many of the authors above the command to
%% show. For example "\collaboration[all]{(DELVE Collaboration)}" wil include
%% all the authors above this command.
%%
%% Mark off the abstract in the ``abstract'' environment. 
\begin{abstract}
Among thousands of extragalactic $\gamma$-ray emitters, only a handful of distant ($z >$ 3) sources are detected. Yet, they are crucial probes shedding light on the cosmic evolution of jets of active galactic nuclei and the initial phase of mass growth of supermassive black holes. Here, we report on a multi-band study of the radio quasar B3 1239+376 with $z$ = 3.82. By analyzing the Fermi-LAT data, a significant (globally 7.7$\sigma$) $\gamma$-ray source in its direction, with an estimated association probability of 0.91, is observed in a half-year period of 2025. The analysis also reveals the emergence of co-spatial $\gamma$-ray residues in prior epochs. Moreover, the $\gamma$-ray and infrared light curves obtained from WISE and SPHEREx observations are likely correlated, as we observe that the emissions in both bands peak at the same time. The temporal coincidence establishes a firm association relationship between the $\gamma$-ray source and the quasar. Therefore, B3 1239+376 is proposed as the third most distant $\gamma$-ray detected blazar to date. Benefiting from the multi-wavelength observations, broadband spectral energy distributions in different flux states are compiled and reproduced by the classic one-zone leptonic radiation model to investigate the jet properties. Considering the recent brightening in $\gamma$ rays, prompt follow-up observations are encouraged, especially radio interferometry observations which may catch the potential ejection of a new jet blob. 

\end{abstract}

%% Keywords should appear after the \end{abstract} command. 
%% The AAS Journals now uses Unified Astronomy Thesaurus (UAT) concepts:
%% https://astrothesaurus.org
%% You will be asked to selected these concepts during the submission process
%% but this old "keyword" functionality is maintained in case authors want
%% to include these concepts in their preprints.
%%
%% You can use the \uat command to link your UAT concepts back its source.

%% From the front matter, we move on to the body of the paper.
%% Sections are demarcated by \section and \subsection, respectively.
%% Observe the use of the LaTeX \label
%% command after the \subsection to give a symbolic KEY to the
%% subsection for cross-referencing in a \ref command.
%% You can use LaTeX's \ref and \label commands to keep track of
%% cross-references to sections, equations, tables, and figures.
%% That way, if you change the order of any elements, LaTeX will
%% automatically renumber them.
\setcounter{footnote}{0}
\section{Introduction} \label{sec:intro}
Active galactic nuclei (AGNs), are one of the most luminous sources of electromagnetic emission in the universe, powered by central supermassive black holes (SMBHs, \citealt{1993ARA&A..31..473A,1995PASP..107..803U}). In the past 10 years, more than 400 quasars with z $>$ 5.7 have been discovered \citep{2023ARA&A..61..373F}. The existence of these objects suggests a prompt and eﬃcient mass growth for their SMBHs, challenging the scenarios of SMBH formation in the early universe \citep{2020ARA&A..58...27I}. The situation becomes more interesting when the high-$z$ AGNs with strong jets (i.e., jetted AGN) are detected \citep[e.g.,][]{2018ApJ...856..105A,2020A&A...635L...7B,2022A&A...663A.147S,2025NatAs...9..293B}. Jetted AGN usually harbor very heavy SMBHs, even in the early cosmic time \citep{2010MNRAS.405..387G,2011MNRAS.416..216V}. The presence of relativistic jets is believed to relate with highly spinning accreting black holes \citep[e.g.,][]{2011MNRAS.418L..79T}, for which a high radiation efficiency is expected ($\eta \sim 0.3$, \citealt{1974ApJ...191..507T}). Therefore, a longer growth time for these SMBHs is needed compared for those residing in non-jetted AGNs, which indicates that the jets might play an important role in the initial phase of SMBH formation \citep{2008MNRAS.386..989J}.

Blazars, which further distinguish into flat-spectrum radio quasars (FSRQs) and BL Lacertae objects (BL Lacs), are a subclass of AGNs for which their strong jets are aligned with our line of sight and hence the beamed non-thermal jet emission dominates the electromagnetic spectrum \citep{2019ARA&A..57..467B}. A distinguished feature of the blazar radiation is the violent variability across diverse timescales \citep[e.g.,][]{1997ARA&A..35..445U}. The broadband jet emissions span from radio to very-high-energy $\gamma$-ray domains, characterized by a universal two-bump spectral shape in log$\rm \nu F_{\nu}$–log$\rm \nu$ plot. One bump is widely accepted from synchrotron emission, while the other one peaking in $\gamma$ rays, is usually explained as inverse Compton (IC) by a population of relativistic electrons, which are also responsible for the synchrotron emission, scattering of soft photons from either inside (the synchrotron self-Compton, or SSC, \citealt{1992ApJ...397L...5M}) and/or outside (external Compton, or EC, \citealt{1993ApJ...416..458D,1994ApJ...421..153S,2000ApJ...545..107B}) the jet. Alternatively, hadronic interactions, for instance the photo-hadronic ($p\gamma$) and hadronuclear ($pp$) processes, are also raised to describe the high energy radiation of blazars \citep{1993A&A...269...67M,2001APh....15..121M,2003ApJ...586...79A}. In addition to the electromagnetic radiation, decay of the pions generated from hadronic processes leads to the neutrino production as well. It is supported by recent multi-messenger observations that spatial and temporal coincidences between the incoming neutrino and the flaring blazars are occasionally detected \citep[e.g.,][]{2018Sci...361.1378I,2025arXiv250208484K}. 

In the $\gamma$-ray sky, the majority of the extragalactic sources are blazars \citep{2023arXiv230712546B}. Although the most distant blazar detected so far is with redshift as high as $z$ = 7 \citep{2025NatAs...9..293B}, the redshift record of $\gamma$-ray-emitting blazars is currently 4.72 (i.e., B3 1428+422, \citealt{2018ApJ...865L..17L,2020ApJ...903..128K,2025ApJ...990..206G}). Actually, there are only two blazars with $\gamma$-ray detections beyond redshift of 4, the other one is GB 1508+5714 ($z$ = 4.3) \citep{2017ApJ...837L...5A,2020ApJ...898L..56L,2024ApJ...974...38G}. $\gamma$-ray radiation from AGN jets, exclusively produced by the non-thermal processes of the relativistic particles, possess a tight connection to the ultra-high-energy cosmic rays \citep{2016ARA&A..54..725M}. Meanwhile, a large fraction of the electromagnetic radiation of blazars, especially the high-$z$ blazars, are released in the $\gamma$-ray domain \citep{2011MNRAS.411..901G}. Moreover, $\gamma$-ray photons from high-$z$ blazars, are a powerful tool for probing the extragalactic background light (EBL, \citealt{2018Sci...362.1031F}). Therefore, the $\gamma$ rays bring key information on the AGN jets such as the acceleration and energy dissipation processes of emitting particles, and shed lights on the cosmic star formation history. However, since these $\gamma$-ray sources are spectrally soft \citep{2017ApJ...837L...5A,2018ApJ...853..159L} and the angular resolution of Fermi-LAT for the sub-GeV photons is worse than that of GeV photons, the major obstacle to detecting $\gamma$-ray signals from high-$z$ blazars is the limited localization. Catching correlated multi-wavelength activities between the $\gamma$-ray emission and those in other windows of electromagnetic radiation acts as a decisive evidence to determine the association relationship \citep[e.g.,][]{2019ApJ...879L...9L}.

B3 1239+376 is first known as a radio source observed by the Bologna Northern Cross Telescope \citep{1973A&AS...11..291C}. It possesses a flat radio spectrum between 1.4 and 5 GHz \citep{1996ApJS..107...37T} and is recognized as a high-$z$ quasar ($z$ = 3.82, \citealt{1996AJ....111.1013V}), which makes it one of the brightest radio quasars in the early universe (i.e., $ f_{5~GHz} \simeq$ 0.5 Jy). Although B3 1239+376 is not listed in the Roma-BZCat (hereafter BZCAT, \citealt{2009A&A...495..691M}), the super-luminous motion of $\beta_{app} \simeq$ 5 has been detected \citep{2008A&A...484..119B}. Meanwhile, its radio-loudness (RL=$f_{1.4~GHz}/f_{g~band}$, \citealt{1989AJ.....98.1195K}) is calculated as high as $\sim$ 25,000, suggesting a blazar nature. In the X-ray perspective, a hard X-ray spectrum (i.e., $\Gamma_{0.5-7~keV}$  = 1.5) observed by Chandra X-ray observatory is suggested, further supporting it as a blazar \citep{2016ApJ...833..123M}. In this study, multiwavelength data of B3 1239+376 are collected and analyzed, and its broadband temporal properties are reported (Section \ref{sec:data}), along with some discussions (Section \ref{sec:discu}). Here we consider a $\Lambda$CDM cosmology with $H_{0}=67~{\rm km~ s^{-1}~Mpc^{-1}}$, $\Omega_{\rm m}=0.32$, and $\Omega_{\Lambda}=0.68$ \citep{2014A&A...571A..16P}.

\section{Data Analysis and Results} \label{sec:data}
\subsection{Fermi-LAT Data} \label{subsec:2-1}
We collected Fermi-LAT Pass 8 {\tt SOURCE} data, with an energy range between 100 MeV and 500 GeV. The data time range is from 2008 August 4 to 2025 August 4  (i.e., MJD 54683 - 60892),  covering a total of 17 years. During the analysis, we adopted the {\tt Fermitools} software version 2.2.0 as well as {\tt Fermitools-data} version 0.18. A zenith angle cut (i.e., $<$ 90$^{\circ}$) along with the recommended quality-filter cuts (i.e., {\tt DATA\_QUAL==1 \&\& LAT\_CONFIG==1}) were included in the data filtering procedures. In the Unbinned likelihood analyses, in which $\gamma$-ray flux and spectrum were extracted by the {\tt gtlike} task, we used the test statistic (TS = -2ln ($ L_{0}$/$L$), \citealt{1996ApJ...461..396M}) to estimate the the significance of the $\gamma$-ray sources. By definition, $L$ and $L_0$ denote the maximum likelihood values for the models with and without the target $\gamma$-ray source, respectively. Since the target is not included in any version of the Fermi-LAT catalogs, including the most recent incremental version (i.e., data release 4 of the fourth Fermi-LAT catalog, or 4FGL-DR4, \citealt{2023arXiv230712546B}), we added a test $\gamma$-ray source at the radio position of B3 1239+376 in the source model. We chosen power-law function to describe the spectral distribution of the test source, of which the initial spectral index was set as 2.5. During the likelihood analysis, all 4FGL-DR4 sources within 15$^{\circ}$ of the target were considered. Spectral and normalization parameters of the sources lying within 10$^{\circ}$ radius of interest (ROI), together with normalization parameters of the diffuse emission templates (i.e., {\tt gll\_iem\_v07} and {\tt iso\_P8R3\_SOURCE\_V3\_v1}), were set free. On the other hand, for sources outside the ROI, we frozen their parameters as the default 4FGL-DR4 values. In the temporal analysis, if detections of the background sources are marginal (i.e., TS $<$ 10), we removed them from the source model. In this case, for the target, 95\% confidence level (CL) upper limits were calculated by the {\tt pyLikelihood UpperLimits} tool, to replace the fluxes. Here we took a linear binning approach, in other words, length of each time bin is equal. Since the variability timescales of blazars are diverse, across minutes to years, different binning strategies were adopted corresponding to different cases. The details were introduced in the following context.

\begin{figure*}
    \centering
    \includegraphics[scale=0.8]{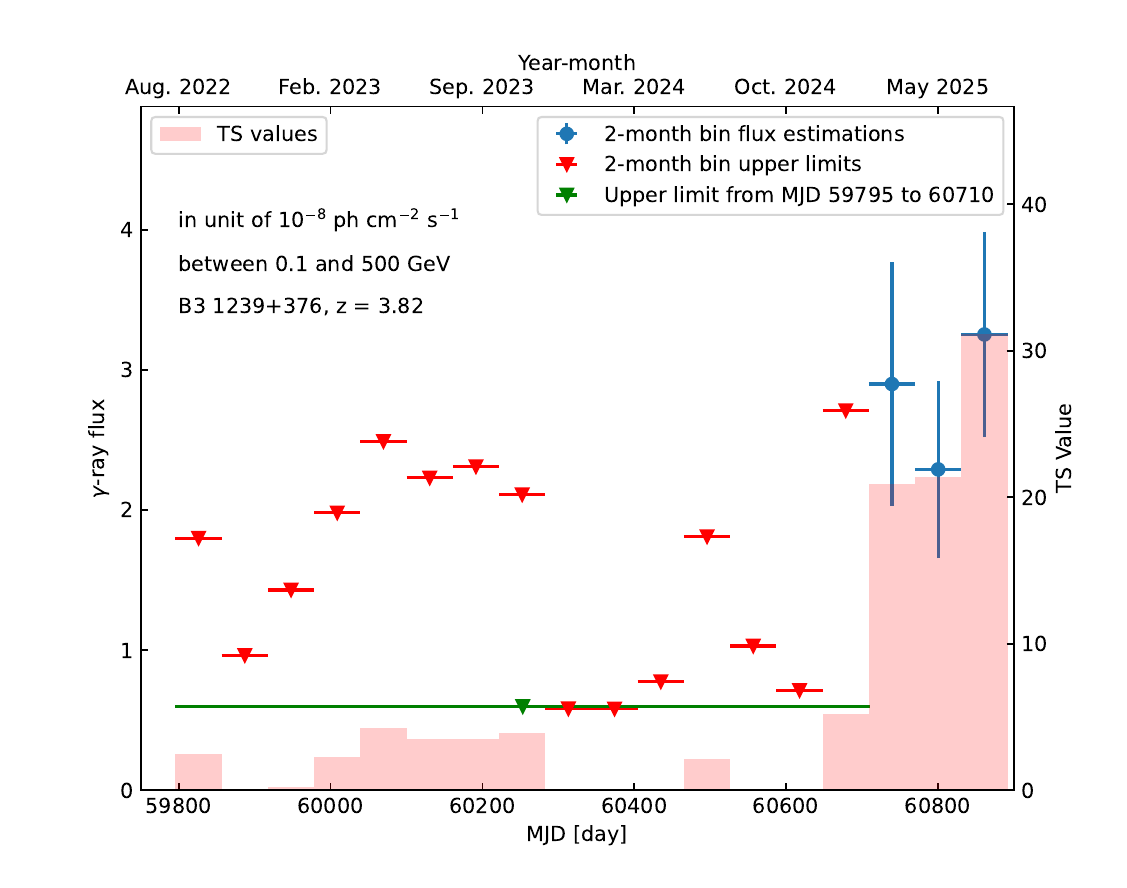}
    \caption{The Fermi-LAT $\gamma$-ray light curve of the source B3~1239+376 between August 2022 and August 2025, with each bin corresponding to a 2-month period. The blue circles represent flux estimates with $1\sigma$ statistical errors, while the red downward triangles are the upper limits. The pink shaded bars show the TS values for each bin. The green horizontal line indicates the averaged flux upper limit for the first 15 bins.
}
    \label{gmlc}
\end{figure*}

\begin{figure*}
\centering
\subfigure[]{
\includegraphics[scale=0.43]{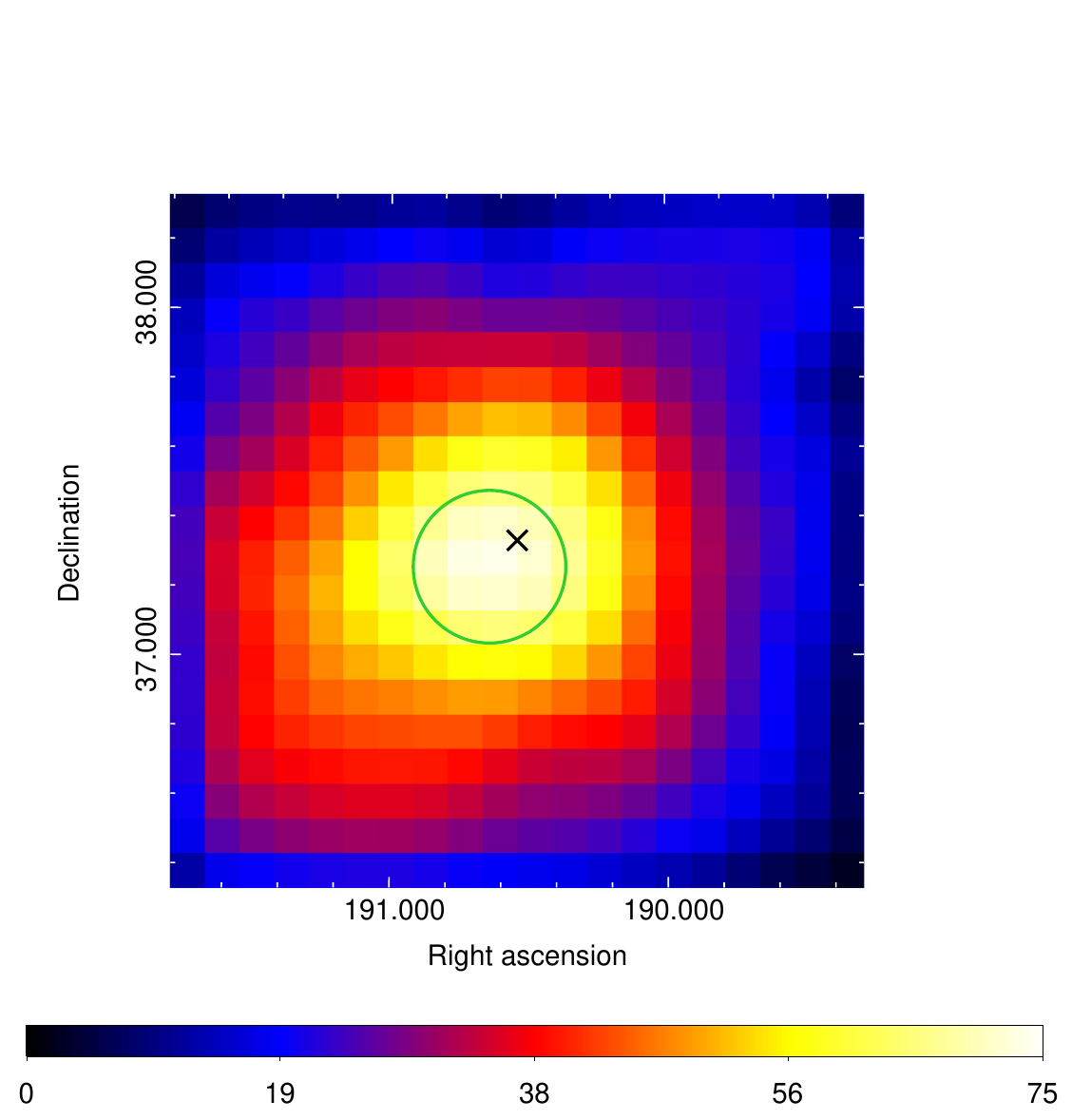}
\label{fmap}}
\subfigure[]{
\includegraphics[scale=0.43]{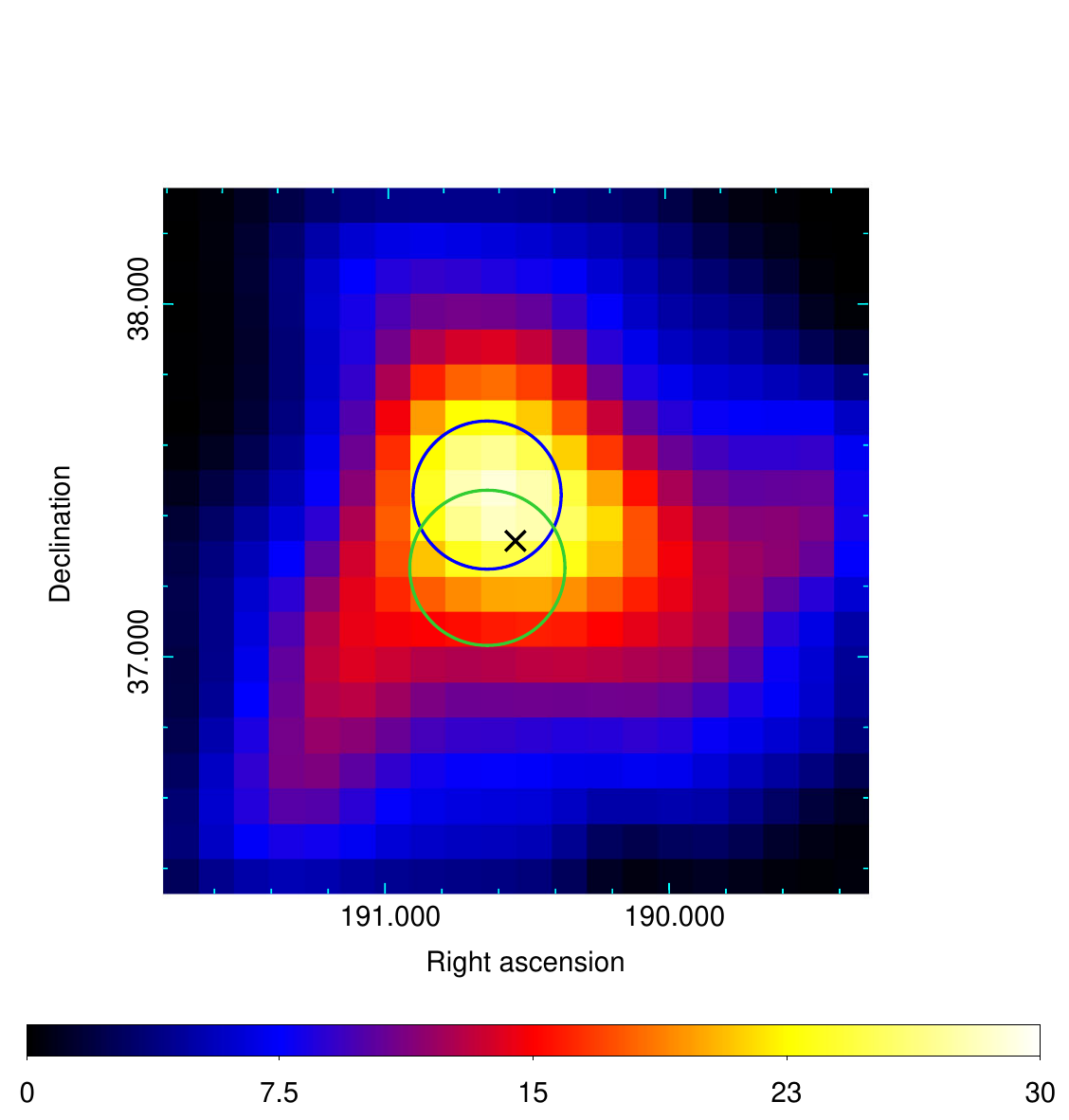}
\label{pfjmap}}
\caption{$\gamma$-ray TS maps centered in the radio position of B3~1239+376 (shown as a black x-shape symbol) that the target source is not included in the analysis source model. They are in a $2^\circ \times 2^\circ$ scale, with an each pixel size of $0.1^\circ$. {\bf Panel~(a)}: the TS map are derived by analyzing Fermi-LAT data between Feb. 2025 and Aug. 2025, from which the 95\% C.L. localization uncertainty is also plotted by the green solid line. {\bf Panel~(b)}: The TS map is based by a joint-likelihood analysis combining two time intervals, from Sep. 2017 to May 2018 and from Nov. 2019 to May 2020, respectively. The corresponding 95\% C.L. localization uncertainty of the joint analysis is drawn in the blue solid line.}
\end{figure*}

Since data from the first 14-yr Fermi-LAT survey (i.e., from 2008-08-04 to 2022-08-02) had been already analyzed and the results were released in 4FGL-DR4 \citep{2023arXiv230712546B}, at first, we focused on data from the recent observations (i.e., from 2022-08-02 to 2025-08-04). Interestingly, an excess in this direction emerged in our analyses, with a TS value of 47 (locally $\simeq$ 5.9$\sigma$). Therefore, a two-month time bin $\gamma$-ray light curve, was extracted to seek any potential activities recently. As shown in Figure \ref{gmlc},  no significant $\gamma$-ray signals were seen until the recent half-year period (i.e., MJD 60710 - 60892). Subsequent analysis aiming this specific half-year data revealed a significant residue (TS = 80, locally $\simeq$ 8.2$\sigma$) towards B3 1239+376, see Figure \ref{fmap}. Here there are 34 trials (i.e., a selected half-year period for the entire 17-year dataset) and hence a global significance, after correction of the trial factor, was estimated as 7.7$\sigma$, suggesting that the detection is statistically significant. Assuming a simple power-law spectral template (i.e. $dN/dE \propto E^{-\Gamma}$, where $\Gamma$ is the photon index), the analysis yields a photon flux of $\rm (3.2\pm0.6)\times10^{-8}$ ph $\rm cm^{-2}$ $\rm s^{-1}$. The source is spectrally soft, with an estimated photon index of $2.60\pm0.14$. By comparison, the fit of the prior 2.5-year data only provided an upper limit (TS = 8),  $\rm 6.0\times10^{-9}$ ph $\rm cm^{-2}$ $\rm s^{-1}$. Therefore, a strong variation with a 5-fold flux increase is indicated. 

Considering the limited angular resolution of Fermi-LAT, we checked the possible influences from the background sources. The closest background source is 4FGL J1243.2+3627, about 0.9$ ^{\circ}$ away, however, it possesses a rather different spectral behavior ($\Gamma \simeq 1.78$, \citealt{2023arXiv230712546B}) than our target. We noted that the brightest background source in the ROI (i.e., 4FGL J1310.5+3221, $\Gamma \simeq$  2.23, \citealt{2023arXiv230712546B}), which is 7.7$^{\circ}$ from the ROI center, underwent a strong outburst in year 2025\footnote[1]{https://fermi.gsfc.nasa.gov/ssc/data/access/lat/LightCurveRepository/source.html?source\_name=4FGL\_J1310.5+3221}. Because the 95\% CL containment angle of Fermi-LAT for 500 MeV photon is $< 6^{\circ}$\footnote[2]{https://www.slac.stanford.edu/exp/glast/groups/canda/lat\_Performance.htm}, by selecting events with energies beyond 500~MeV, we carried out a further analysis. In this case, the contamination from the bright source could be significantly diminished. The detection remains to be significant, with a TS value of 39 which corresponds to a global significance of 4.7$\sigma$. Therefore, the new $\gamma$-ray source is likely intrinsic than artificially caused by the background sources. The obtained photon flux and index from the $>$ 500~MeV data analysis are consistent with that from the full data energy range.

Then we performed the localization analysis of the $\gamma$-ray source. The optimized location was constrained as R.A. 190$^{\circ}$.6411 and decl. 37$^{\circ}$.2587, with a 95\% CL uncertainty radius of 0.22$^{\circ}$. Corresponding angular separation between that and the radio position of B3 1239+376 is 0.11$^{\circ}$, and hence it falls into the $\gamma$-ray localization area, see Figure \ref{fmap}. When low energy threshold of the photons was set to 500~MeV, the best location is R.A. 190$^{\circ}$.4602 and decl. 37$^{\circ}$.2475. At this time, the angular separation (0.11$^{\circ}$) is still smaller than the 95\% CL uncertainty radius (0.21$^{\circ}$). We also searched other potential low-energy association candidates. Since blazars are the dominant population of the extragalactic sky, we firstly looked into the blazar candidate catalogs. No co-spatial sources were found in either the BZCAT list \citep{2009A&A...495..691M} were found nor the WIBRaLS2 catalog \citep{2019ApJS..242....4D} was investigated. Additionally, we investigated low-energy counterparts for $\gamma$-ray source by searching the high-frequency radio survey as well as from the VLBI observations \citep{2020ApJS..247...33A}. No other sources than B3 1239+376 in the CRATES catalog were found \citep{2007ApJS..171...61H}, so as the radio fundamental catalog \citep{2025ApJS..276...38P} in this direction. Adopting the Bayesian association method together with the corresponding prior probability value for CRATES catalog (0.33, \citealt{2010ApJS..188..405A}), we calculated the association probability as 0.91, above the threshold value of 0.8. Hence, B3 1239+376 is suggested as the counterpart to the $\gamma$-ray source. Nevertheless, considering the relatively large localization uncertainty in the $\gamma$ rays, detections of multiwavelength correlated emissions are needed to establish a firm association relationship.

To provide a global picture of the temporal behavior of the $\gamma$-ray source, we built a 1-year time bin light curve for the entire 17-year dataset, as shown in Figure \ref{multi-lc}. Adopting the ``variability index" test \citep{2012ApJS..199...31N}, the variability is suggested to be significant ($5.4\sigma$). The recent brightening (i.e., the last time bin) stands out, and it has been picked up by the Bayesian block approach \citep{2013ApJ...764..167S} indicative of a flare event, which is consistent with the results mentioned above. At the initial 9-year Fermi-LAT observation, the target was in a low flux state. The corresponding analysis yielded a marginal detection (TS =12), with a photon flux constrained to $\rm (2.4\pm0.3)\times10^{-9}$ ph $\rm cm^{-2}$ $\rm s^{-1}$ by setting the photon index as 2.6.  It is noteworthy that there are tentative detections (TS $\gtrsim$ 15) in two yearly time bins around year 2017 and 2020, respectively. Then a 4-year long quiescent state was followed, from which an upper limit of $\rm 3.8\times10^{-9}$ ph $\rm cm^{-2}$ $\rm s^{-1}$ was obtained. 

We carried out detailed investigations on the potential signals from year 2017 to 2020. In the first step, we extracted a 1-month time bin light curve, to identify the exact time range of the signals. We selected two epochs, in which one is a 6-month period between MJD 58060 and 58242, and the other one lasting for 10 months is from MJD 58727 to 58969. The corresponding TS values are 15 and 17, respectively. Since analyzing Fermi-LAT data in neither of the epochs alone yield significant detections (i.e., TS $>$ 25), we performed a joint analysis of the two epochs then. The joint analysis returned an increased TS value (TS = 31) and the enhanced statistics allowed us to localize the residue. The optimized location was given as R.A. 190$^{\circ}$.642 and decl. 37$^{\circ}$.4649, along with a 95\% CL uncertainty radius of 0.21$^{\circ}$. As shown in Figure \ref{pfjmap}, B3 1239+376 is included within the location error radius, since the angular separation is 0.15$^{\circ}$ under this circumstance. We also derived the photon flux and the spectral information. The former was obtained as $\rm (1.3\pm0.4)\times10^{-8}$ ph $\rm cm^{-2}$ $\rm s^{-1}$, while the photon index was constrained as $\rm 2.59\pm0.19$ that is in agreement with the result from the flare in year 2025. Both the spatial and spectral information extracted in the joint analysis suggest the potential activities then likely also related to B3 1239+376.

\subsection{X-ray data} \label{subsec:2-2}
\subsubsection{Chandra data} \label{subsec:2-2-1}
B3 1239+376 was observed twice by the $Chandra$ Advanced CCD Imaging Spectrometer (ACIS, \citealt{2003SPIE.4851...28G}), on 2007 Mar. 10 (ACIS-7, ObsID: 7871, PI.: C. Cheung) and 2022 Dec. 31 (ACIS-67, ObsID 26906; PI: E. Meyer). The corresponding exposure times are 5.2 and 20.1 ks, respectively. The observations were performed on the pointing mode, events were recorded in the  {\tt Faint} telemetry format and with the  {\tt Timed} Exposure mode. We carried out the data reductions using the Chandra Interactive Analysis of Observations software package \citep{2006SPIE.6270E..1VF} v4.17 with {\tt CALDB} version 4.12. Initially, we reprocessed the standard level 1 event lists into the level 2 data products by the {\tt chandra\_repro} script, in which standard grade, status, and good time filters, as well as {\tt VFAINT} background cleaning were applied. Meanwhile, we removed background flaring periods by the routine {\tt deflare}. We estimated the possible pileup effect using PIMMS version 4.15\footnote{https://heasarc.gsfc.nasa.gov/cgi-bin/Tools/w3pimms/w3pimms.pl}. Since B3 1239+376 had $< 5$\% pileup over the 0.5–7.0 keV energy band, pileup was ignored for this study.

\begin{figure*}
    \centering
    \includegraphics[scale=0.8]{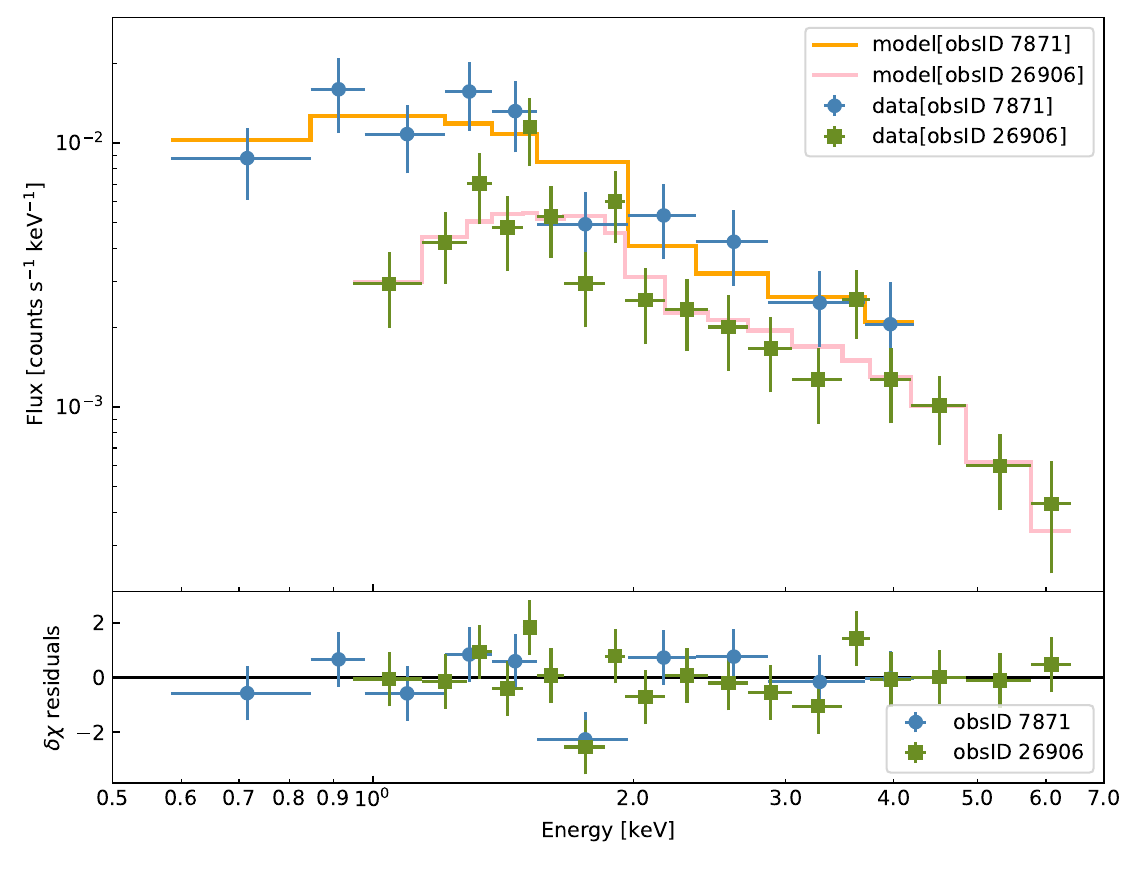}
    \caption{The folded X-ray spectra of B3~1239+376 from two \emph{Chandra} observations (top panel), as well as fit residuals (lower panel). The solid lines in the top panel represents the best-fit models. Note that due to the limited exposure time the most energetic X-ray photons for the observation in year 2007 (data points in color blue) are at $\simeq$ 4~keV. To conveniently compare with the flux derived from the other observation, for the both observations we took a routinely used energy range (i.e., 0.5-7 keV) to calculate the flux measurements.}
    \label{chandra}
\end{figure*}

We used the \texttt{specextract} task in the X-ray spectral analysis. We extracted the target events from a circular region with a radius of $4\arcsec$ centered at the position of B3 1239+376, while the background spectrum was extracted from two circular regions with radius $25\arcsec$ in blank sky regions. The numbers of observed net photons of the observation in year 2007 and 2022 are 117 and 198, respectively. We grouped the spectra to have at least five counts per bin. Then we used the {\tt XSPEC} tool (\citealt{1996ASPC..101...17A}, version 12.14.0h) to perform the spectral fitting procedure, in which the energy range was set as 0.5 -- 7~keV and the $\mathcal{C}$-statistic \citep{1979ApJ...228..939C} was adopted. The spectral model was assumed as a power-law distribution absorbed by the Galactic column density along the line of sight ($N_{H}$ = $\rm 1.25\times10^{20}~cm^{-2}$, \citealt{2016A&A...594A.116H}). The data together with the optimized models were shown in Figure \ref{chandra}. The analysis on the observation in year 2007 gave an unabsorbed flux of $f_{0.5-7 keV}=2.4^{+0.74}_{-0.55}\times10^{-13}\ {\rm erg\ cm^{-2}\ s^{-1}}$ ($\mathcal{C}$-Statistic/d.o.f., 12.3/17), along with a hard spectrum $\Gamma_{x} = 1.14\pm0.33$. While the results on the other observation indicates a possibly fainter emission, $f_{0.5-7 keV}=1.8^{+0.22}_{-0.21}\times10^{-13}\ {\rm erg\ cm^{-2}\ s^{-1}}$ ($\mathcal{C}$-Statistic/d.o.f., 26.3/33). The X-ray emission in year 2022 is confirmed to be spectrally hard, $\Gamma_{x} = 1.51\pm0.25$.

\subsubsection{Swift data} \label{subsec:2-2-1}
There are six snapshots from Neil Gehrels Swift Observatory \citep{2004ApJ...611.1005G} in the direction of B3 1239+376. The observational target is RX 1241.4+3722, a nearby Seyfert I galaxy roughly 8.3\arcmin~away from the high-$z$ quasar. We analyzed the XRT photon counting mode data as well as the UVOT data with the FTOOLS software version 6.33.2. Initially, we performed the event cleaning for the XRT data by the {\tt xrtpipeline} task using standard quality cuts. Then we utilized the {\tt ximage} task to search any X-ray sources in the field. Unfortunately, no significant (i.e., signal-to-noise, S/N $>$ 3) excesses over the background towards B3 1239+376 were found. For the UVOT images, we carried out aperture photometries by the {\tt uvotsource} task, with a 5 arcsec circular aperture along with a background extraction in a larger source-free region. One marginal detection (S/N = 3.4, 21.02 $\pm$ 0.02 mag), in B-band at MJD 56220, was revealed.

\subsection{SDSS spectrum}\label{subsec:2-2}
\begin{figure*}
    \centering
    \includegraphics[scale=0.8]{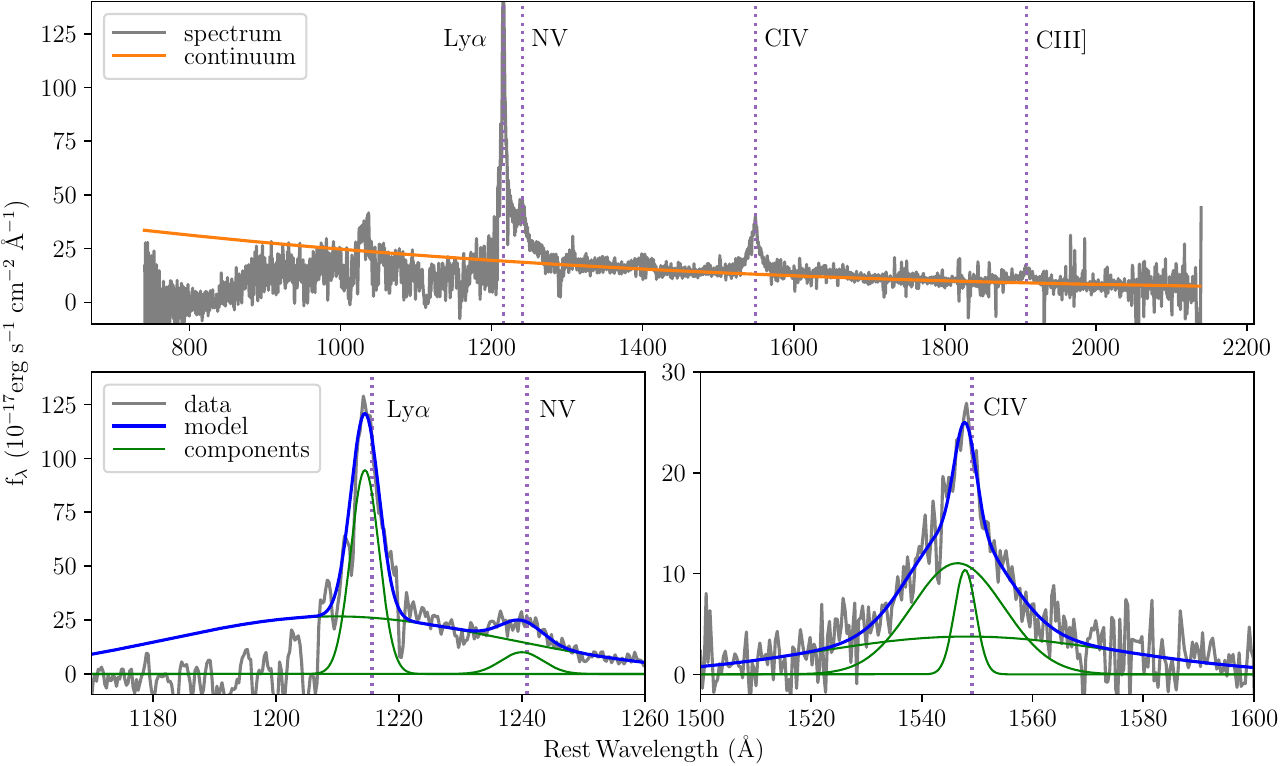}
    \caption{Fitting of the SDSS spectrum of B3~1239+376. A global scope of the spectrum (in grey) and the modeled continuum component (in orange) are shown in the upper panel. The distinct emission lines are marked as well. In the lower panel, zoomed-in views on the Ly$\alpha$+N\,\textsc{v} (left) and C\,\textsc{iv} (right) line regions are provided especially.}
    \label{sdss}
\end{figure*}

During the fit of the SDSS spectrum from SDSS data release 16 \citep{2020ApJS..249....3A}, we firstly removed effects of galactic extinction, and described the continuum by a power law and polynomial model plus the Fe \textsc{ii} template \citep{2001ApJS..134....1V,2006ApJ...641..689V}. We derived the uncertainties from 1000 mock spectra, generated by adding Gaussian noise consistent with the measurement errors. The standard deviation of the best-fit parameters across these realizations provided the uncertainty for each spectral quantity. As shown in Figure \ref{sdss}, several emission lines are distinct, especially the Ly$\alpha$ and the \civ\ lines. The redshift is constrained as $z$ =3.82 $\pm$ 0.01, consistent with the value listed in SDSS quasar catalog DR16 \citep{2022ApJS..263...42W}. For the Ly$\alpha$ emission line region, we used two components to fit the Ly$\alpha$ line and one component to fit the N\textsc{V} line. The $L_{\rm Ly\alpha}$ is obtained as $(1.52\pm0.03) \times10^{45}$ $\rm erg~s^{-1}$, while the $L_{1350\,\text{\AA}}$ is $(3.06\pm0.04) \times10^{46}$ $\rm erg~s^{-1}$. Taking the bolometric correction factor $\rm BC_{1350}$ = 3.81 \citep{2011ApJS..194...45S}, the bolometric disk luminosity $\rm L_{bol,disk}$ is suggested as high as $\simeq$ $10^{47}$ $\rm erg~s^{-1}$. As for the \civ\ line fitting, three Gaussian components, including one narrow component and two broad components, were adopted. We estimated the $\rm FWHM_{C~IV}$ as $2112 \pm 252\ \mathrm{km\ s^{-1}}$. However, note that the \civ\ line is known to be biased by strong outflows, and hence one should be cautious to base black hole mass estimations on this emission line \citep{2023ApJ...950...96D}. Future near-infrared spectroscopy by a ground-based telescope, like Keck/NIRES \citep{2004SPIE.5492.1295W}, would give a robust BH mass estimation of B3 1239+376 \citep{2024MNRAS.527.5356B}.

\subsection{Infrared data}\label{subsec:2-3}
\begin{figure*}
    \centering
    \includegraphics[scale=0.8]{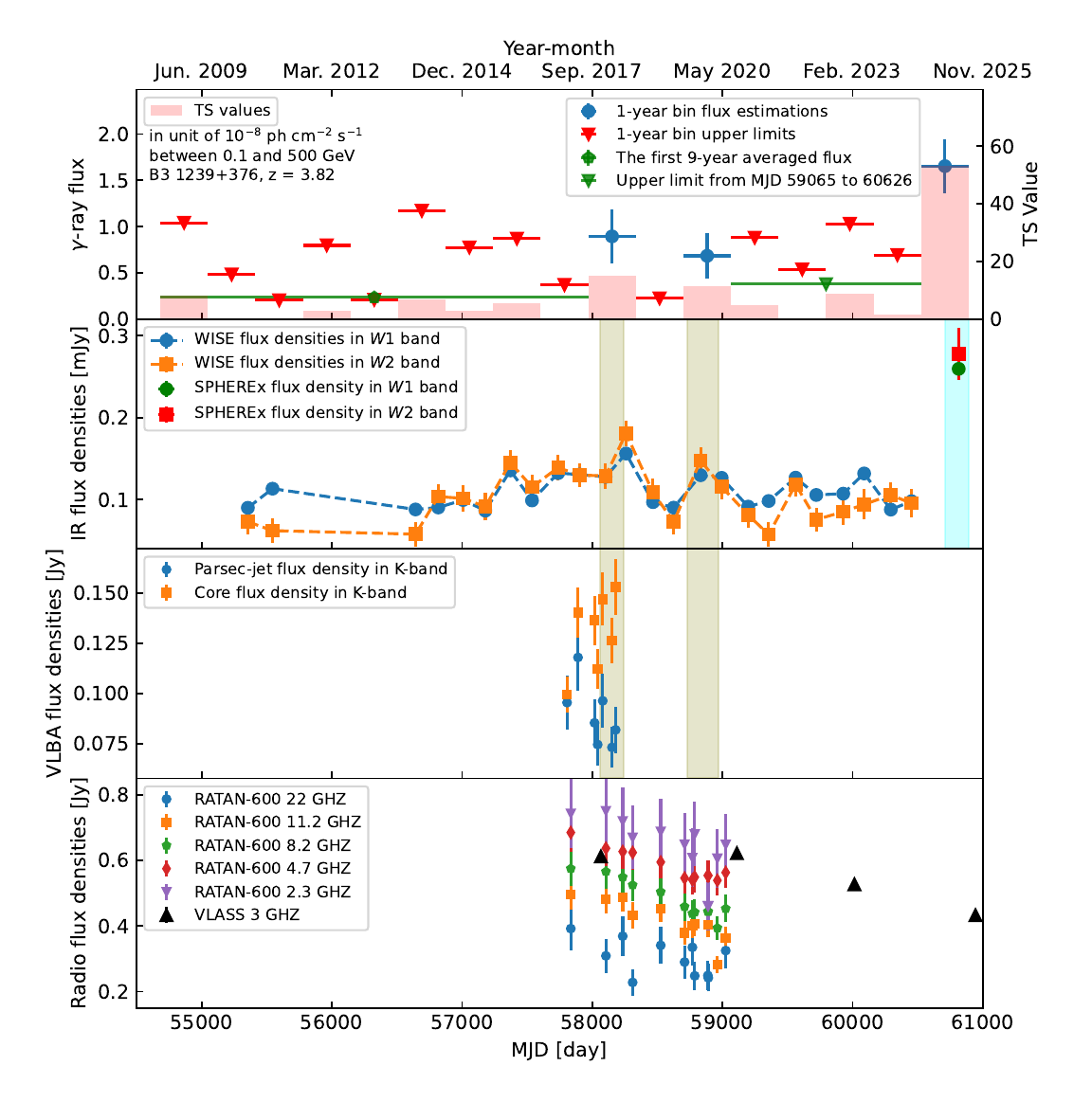}
    \caption{Multi-wavelength long-term light curves of B3 1239+376. {\bf Top panel}: Fermi-LAT 1-year time bin $\gamma$-ray light curve. Blue points denote flux measurements, red triangles are 95\% CL upper limits, and the pink histogram shows the corresponding TS values. The green horizontal line marks the averaged flux over the first 9-year Fermi-LAT observation, and the green triangle is the upper limit for a four-year quiescent state. {\bf Upper middle panel}: Infrared light curves in the W1 ($3.4~\mu\mathrm{m}$) and W2 ($4.6~\mu\mathrm{m}$) bands, observed by WISE/NEOWISE and SPHEREx. The vertical filled regions in color olive marks the time ranges for the joint Fermi-LAT analysis, of which the TS map is presented in Figure \ref{pfjmap}.  While the filled region in color cyan represents the half-year $\gamma$-ray flaring period in 2025, corresponding to the TS map in Figure \ref{fmap}. {\bf Lower middle panel}: VLBA K-band ($24~\mathrm{GHz}$) flux densities of the parsec-scale jet and compact core components, shown in blue circles and orange squares, respectively. {\bf Bottom panel}: Radio light curves from RATAN-600 at 2.3–22 GHz together with VLASS 3 GHz flux measurements.}
    \label{multi-lc}
\end{figure*}

B3 1239+376 is included in the AllWISE Source Catalog \citep{2010AJ....140.1868W,allwise}, named as WISEA J124209.81+372006.0. The \emph{W1} ($3.4~\mu$m) and \emph{W2}($4.6~\mu$m) magnitudes listed in the catalog are 16.22 $\pm$ 0.06 and 15.90 $\pm$ 0.15, respectively. The detections in \emph{W3} ($11~\mu$m) and \emph{W4}($22~\mu$m) bands are not statistically significant. We extracted long-term infrared light curves in \emph{W1} and \emph{W2} bands , based on the time-resolved WISE/NEOWISE co-added images~\citep{2018AJ....156...69M}, due to the faint infrared radiation. During the analyses, we carried out point spread function photometry, targeting the difference images that the first epoch observations were treated as reference images \citep{2021ApJ...911...31J}. Along with the long-term $\gamma$-ray light curve, the infrared light curves are also shown in Figure \ref{multi-lc}. We utilized a $\chi^{2}$-test with a constant flux level as the null hypothesis, to investigate the significance of the infrared variations. The minimum reduced $\chi^{2}$ values obtained by varying the constant flux levels in \emph{W1} and \emph{W2} light curves are 7.2 and 3.9 (d.o.f. = 23), respectively. Therefore, the infrared variability is suggested to be significant ($> 5 \sigma$). The variation tendency in the two bands are similar. The variability amplitude in \emph{W2} band appears to be higher than that in the \emph{W1} band, but the former also has larger measure uncertainties. At the start of the WISE survey, B3 1239+376 was in the quiescent state, after year 2014, a flux rise was apparent lasting about $\sim$ 1500 days. The detected flux maxima were at MJD 58260, after which a flux decline followed. The emissions bounced back with a peak at MJD 58832. There is a minor flux activity later around MJD 59562.

Although the WISE mission was ended on 31th July 2024 (i.e., MJD 60522), a new infrared space telescope, named as Spectro-Photometer for the History of the Universe, Epoch of Reionization and Ices Explorer (SPHEREx, \citealt{2025arXiv251102985B}), has been successfully launched in Mar. 2025. It performs a 0.75-5 \micron~spectroscopy survey over the entire sky. Based on its first reprocessed quick release products \citep{SPHERExQR2}, we obtained one spectrum on B3 1239+376 observed from MJD 60800 to MJD 60829. To compare with the WISE measurements, we convolved the SPHEREx spectrum with the WISE \emph{W1} and \emph{W2} filter transmission curves\footnote{https://svo2.cab.inta-csic.es/svo/theory/fps3/index.php}. We derived the SPHEREx \emph{W1} and \emph{W2} flux densities as (0.259 $\pm$ 0.008) mJy and (0.278 $\pm$ 0.032) mJy, respectively, see Figure \ref{multi-lc}. Interestingly, these values are unprecedentedly high, roughly 2-3 times higher than the average fluxes observed by WISE.

\subsection{Radio data}\label{subsec:2-4}
B3 1239+376 is a well studied radio source. It was monitored by RATAN-600 \citep{1993IAPM...35....7P}, in 12 epochs between MJD 57836 and 59026, at multi-bands from 1.2~GHz to 22~GHz \citep{2021MNRAS.508.2798S}. In addition, seven snapshots at K-band (24~GHz) of Very Long Baseline Array (VLBA) observations (from year 2017 to 2018) with submilliarcsecond resolution were carried out \citep{2023AJ....165..139D}. We looked into images from the Very Large Array Sky Survey (VLASS, \citealt{2020PASP..132c5001L}), from which radio fluxes at 3~GHz were extracted by modelling the source with a Gaussian fit on the image plane using the Common Astronomy Software Applications. We derived the VLASS flux density in the first epoch as 612 $\pm$ 9 mJy, which is consistent with the value (i. e., 608 $\pm$ 3 mJy) in epoch 1 quick look catalog of VLASS \citep{2020RNAAS...4..175G}. The radio data are also plotted in Figure \ref{multi-lc}. The RATAN-600 multi-bands light curves exhibit a general behavior of flux decay in long-timescale, which is supported by the results of the VLASS observations. Two components, including a compact core as well as a parsec-scale jet, are resolved in the VLBA images, interestingly, from which different temporal behaviors are revealed. For the core, the K-band flux density is 0.099 Jy at MJD 57807 and then reaches to 0.153 Jy at MJD 58180, with an average relative uncertainty of 9\%, indicating a likely flux increase. On the other hand, no significant variation of the jet flux density is found. 

\subsection{Implications from the multi-wavelength observations}\label{subsec:2-4}
Through a detailed investigation of the multi-wavelength properties of B3 1239+376, its nature as a high-$z$ $\gamma$-ray-emitting blazar is revealed. In addition to its typical blazar behavior recorded in literature, for instance, the extremely high radio loudness, super-luminous motion and hard X-ray spectrum \citep{2008A&A...484..119B,2016ApJ...833..123M}, in this study, $\gamma$-ray signals detected by Fermi-LAT toward B3 1239+376 are reported. In year 2025, a robust (7.7$\sigma$ in global) $\gamma$-ray source there emerges, with a relatively soft spectrum ($\Gamma$ = 2.6). The half-year averaged photon flux then is one order of magnitude of the value that we derived from the initial 9-year Fermi-LAT observations, indicating a strong $\gamma$-ray flare. B3 1239+376 is known as the unique blazar (candidate) in the direction and the association probability is calculated as 0.91, which establishes a physical connection between the $\gamma$-ray source and B3 1239+376. Moreover, by revisiting the Fermi-LAT data, prior $\gamma$-ray activities linking to the target are also found. Although these two detections are marginal (TS $\geq$ 15) each alone, a joint analysis brings enhanced statistics (TS = 31) enabling us to perform a localization analysis, from which B3 1239+376 also falls into the localization uncertainty, shown in Figure \ref{pfjmap}. Most importantly, behaviors of the $\gamma$-ray and infrared light curves are rather similar. At a scope of 16-year monitoring, $\gamma$-ray flux reaches up to a maximum level in year 2025, at the same time, the infrared fluxes are unprecedentedly high. In addition, the appearance of the prior tentative $\gamma$-ray signals is temporally coincident with the times of the high flux levels in the WISE light curves, see Figure \ref{multi-lc}. The coincidence acts as the decisive evidence to pin down the association relationship. The IR and $\gamma$-ray observations suggest that in year 2017/2020 B3 1239+376 underwent mild activities while a strong burst came in the first half of year 2025. The most energetic $\gamma$-ray photon relative to B3 1239+376 is at energy of $\simeq$ 1~GeV, which could not challenge the current EBL models \citep[e.g.,][]{2010ApJ...712..238F}. Although the data sampling of the WISE observations is limited, detections of contemporaneous brightening in the Fermi-LAT and infrared light curves play an important role in identification of high-$z$ $\gamma$-ray-emitting blazars \citep{2019ApJ...879L...9L,2020ApJ...900...72L}. It is also worth mentioning that the k-band core flux climbs when the $\gamma$-ray and infrared emissions brighten. Unfortunately, no continual VLBA monitoring after the activity is available and detecting a potential ejection of a new jet blob could not be accomplished then.

\section{SED modeling} \label{sec:sed}
\begin{figure*}
    \centering
    \includegraphics[scale=0.6]{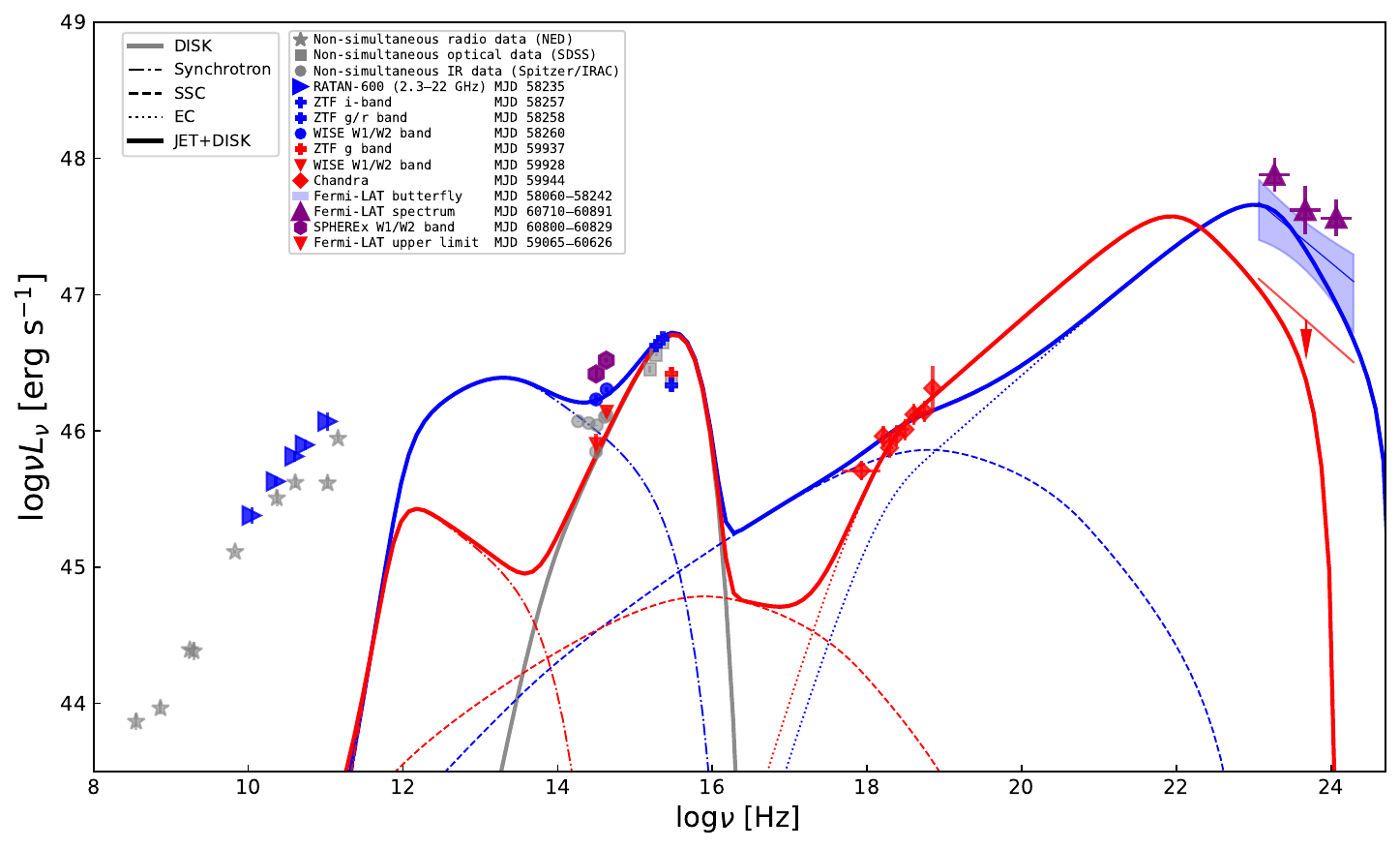}
    \caption{Broadband spectral energy distribution of B3~1239+376. The corresponding input parameters are listed in Table 1. The data and modeling in high-flux state are plotted in blue, while those in the low-flux state are in red. The $\gamma$-ray and infrared data corresponding to the flare in year 2025, are exhibited in color purple. Non-simultaneous data are drawn in grey. The description of the big blue bump is plotted in a solid grey line, $\rm L_{d} \simeq 10^{47}$ erg $\rm s^{-1}$ ($\rm \sim 0.5L_{edd}$) and $\rm M_{BH}$ is set as $\rm 10^{9}M_{\odot}$.
    }
    \label{sed}
\end{figure*}

The multiwavelength observations not only bring valuable temporal information of B3 1239+376, but also allow us to compile the broadband SEDs, presented in Figure \ref{sed}. According to the high flux state, quasi-simultaneous data including the $\gamma$-ray spectrum between MJD 58060 and 58242, ZTF g and r band photometry at MJD 58258 as well as the i band one at MJD 58257, WISE \emph{W1} and \emph{W2} band measurements at MJD 58260, and the RATAN-600 flux densities at MJD 58235 from 2.3 to 22~GHz. On the other hand, the low-flux-state SED consists of the Fermi-LAT upper limit from MJD 59065 to 60626, the Chandra snapshot at MJD 59944, the ZTF g band exposure at MJD 59937, and WISE \emph{W1} and \emph{W2} flux densities at MJD~59928. Additionally, $\gamma$-ray and infrared data corresponding to the flare event in year 2025 are also presented, although they are not modeled due to the limited data coverage. Non-simultaneous archival data are collected and plotted as backgrounds, such as the radio data collected from NED, the optical photometric data from SDSS and multi-bands infrared data from Infrared Array Camera (IRAC) aboard Spitzer. 

Since no significant flux variations are seen in the optical r and g band, the accretion disk emission is likely dominant. The big blue bump is described as a standard \cite{1973A&A....24..337S} disk, extending from 3$R_{s}$ to 2000$R_{s}$, where $R_{s}$ is the Schwarzschild radius. The total disk luminosity is produced as $L_{d}=\eta\dot{M}c^{2}$, in which $\dot{M}$ represents the accretion rate and the accretion efficiency $\eta$ is set as a typical value, 0.1. The accretion disk emission is characterized as a multi-temperature radial profile. At a certain radius $R_{d}$, the local temperature is,
\begin{equation}
T = \left\{\frac{3R_{s}L_{d}}{16\pi\eta\sigma_{SB}R_{d}^{3}}\left[1-(\frac{3R_{s}}{R_{d}})^{1/2}\right]\right\}^{1/4},
\end{equation}   
where $\sigma_{SB}$ is the Stefan-Boltzmann constant. Since the g band is likely drop-out due to neutral intergalactic medium absorption, it is not considered for the modeling of the disk. It is remarkable that the infrared radiations significantly vary for different flux states. In the low flux state, the \emph{W1} and \emph{W2} band data points are in alignment with the accretion disk radiation, while the fluxes increase, especially for the \emph{W2} band, in the high flux state when the jet contribution is un-ignorable. The jet activity in the infrared bands is also supported by the non-simultaneous IRAC observations \citep{2010ASPC..434..437T}. The change of the $\gamma$-ray spectra is also clear. In the 4-year long low flux state, only an upper limit is yielded for Fermi-LAT. While a significant $\gamma$-ray signal is revealed in the half-year period of 2025. Due to the limited statistics, a butterfly plot is drawn for the $\gamma$-ray spectrum of the high-flux-state SED.

Here a classic homogeneous one-zone leptonic radiation scenario is used to give a description of the jet emission. A relativistically moving compact blob with a radius of $\rm R_{j}^{\prime}$ embedded in the magnetic field (B) is responsible for the emission. Within the blob, it is assumed that the emitting electrons follow a broken power-law distribution,
\begin{equation}
N(\gamma )=\left\{ \begin{array}{ll}
                    K\gamma ^{-p_1}  &  \mbox{ $\gamma_{\rm min}\leq \gamma \leq \gamma_{br}$} \\
            K\gamma _{\rm br}^{p_2-p_1} \gamma ^{-p_2}  &  \mbox{ $\gamma _{\rm br}<\gamma\leq\gamma_{\rm max}$,}
           \end{array}
       \right.
\label{Ngamma}
\end{equation}
where the $p_{1,2}$ are indices of the broken power-law distribution, $\rm \gamma _{min}$, $\rm \gamma_{br}$ and $\rm \gamma _{max}$ are the minimum, breaking and maximum energies of the electrons, respectively, and $K$ is the normalization of the particle number density. The transformations of frequency and luminosity between different frames (i.e., the jet comoving frame and the observational frame) are $\nu = \delta\nu^{\prime}/(1+z)$ and $\nu L_{\nu} = \delta^{4}\nu^{\prime}L^{\prime}_{\nu^{\prime}}$, where $\delta$ is the jet Doppler factor. In the leptonic radiation scenario, both synchrotron and IC processes are included. For the former, the synchrotron self-absorption (SSA) process is considered. In the IC processes, the Klein$-$Nishina effect is taken into account. Because of the strong Ly$\alpha$ line (i.e., 2.5$\times10^{15}$ Hz at the AGN frame) detected in the SDSS spectrum, it is adopted as the external soft photons in the EC process. Assuming the broad-line photons are 10\% of the accretion luminosity in a shell-like shape located at a distance $\rm R_{BLR} = 10^{17} L^{1/2}_{disk,45}$ cm \citep[e.g.,][]{2011MNRAS.411..901G}, the corresponding energy density is set as $\rm u_{ext}$ = 2 $\times10^{-2}$ erg $\rm cm^{-3}$. The radius of the emitting blob is usually constrained by the observed variability timescale in short term, $\rm R_{j}^{\prime} \lesssim c \delta \tau_{var}/(1+z)$, however, no such information of B3 1239+376 is available. Thus, a typical value of $\rm R_{j}^{\prime} = 2 \times10^{16}$ cm is used.

Due to the limited data coverage of the SEDs, we adopted a ‘fit-by-eye’ strategy to carry out the SED modeling. For the high-flux-state SED, the radio data points were not considered in the jet modeling, because they are likely from an extended region rather than the compact blob, and the radio emissions are seriously suffered by the SSA. Meanwhile, the optical emissions are dominated by the accretion disk. The remaining data points are insufficient to conduct a $\chi^{2}$-test. While for the low-flux-level SED, the situation is similar. We fixed majority of the input parameters for the SEDs of different flux states, such as $p_{1,2}$ and $\rm \gamma _{min}$, $\rm R_{j}^{\prime}$, as well as $\delta$. We only allowed a change of a minimal amount of parameters (i.e., B, $K$ and $\rm \gamma_{br}$). Nevertheless, the simple leptonic model can reasonably reproduce the both SEDs, see Figure \ref{sed}. The corresponding input parameters are listed in Table \ref{tpara}.

\section{Discussions and Summary} \label{sec:discu}
\begin{figure*}
    \centering
    \includegraphics[scale=0.5]{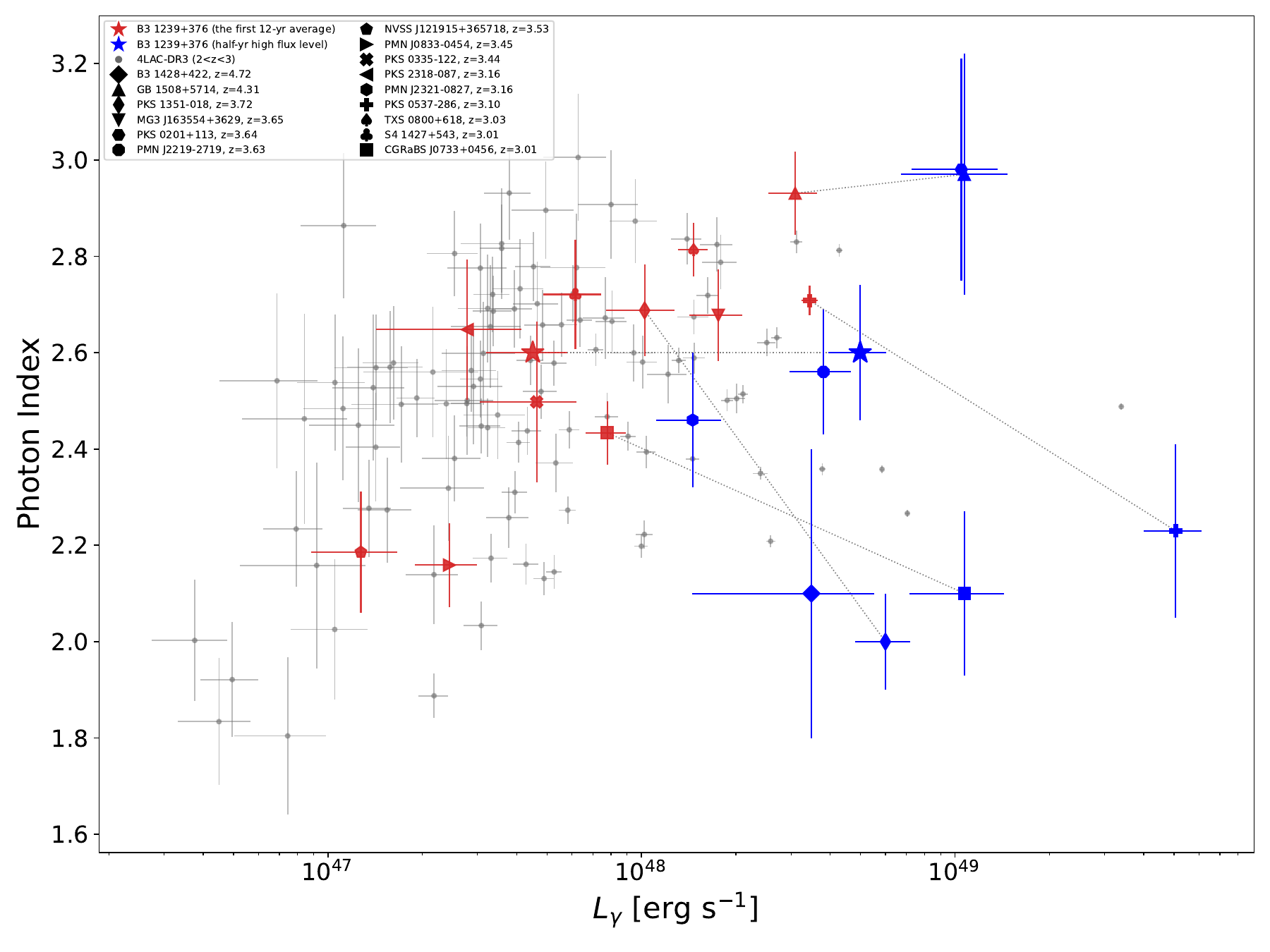}
    \caption{$L_{\gamma}$ vs. $\Gamma_{\gamma}$ diagram for high-redshift blazars. Majority of the luminosities (between 0.1 - 300 GeV) are calculated by the observed energy flux density and $\Gamma_{\gamma}$ listed in 4FGL-DR3 \citep{2022ApJS..263...24A}, in which the $k$-correction is considered. 4LAC-DR3 sources at $2<z<3$ are plotted as backgrounds in grey dots, while red markers represent sources with $z >$ 3. By comparison, a few sources with available information in high-flux state (blue markers) reported in literature are presented as well. B3~1239+376 is plotted in a star shape. The blue star corresponds to the half-year high flux state in 2025, while no reliable spectral information can be obtained for the first 12-year Fermi-LAT observations (i.e., the red star).
    }
    \label{comp}
\end{figure*}

High-redshift (z $>$ 3) $\gamma$-ray blazars are rare. Among the thousands of extragalactic $\gamma$-ray emitters \citep{2022ApJS..263...24A}, only a handful of these sources are found \citep[e.g.,][]{2017ApJ...837L...5A}. Yet they are crucial for our understanding in AGN jet cosmic evolution, the star formation history of the universe, and the initial mass growth of SMBHs. In particular, besides the two sources above $z$ of 4, B3 1428+422 and GB 1508+5714 \citep{2017ApJ...837L...5A,2018ApJ...865L..17L}, there are five known blazars with $z >$ 3.5 detected by Fermi-LAT to date. They are NVSS J121915+365718 ($z$ = 3.53, \citealt{2020ApJ...903L...8P}), PMN J2219-2719 ($z$ = 3.63, \citealt{2020ApJ...900...72L}), PKS 0201+113 ($z$ = 3.64, \citealt{2024ApJ...970..185L}), MG3 J163554+3629 ($z$ = 3.65, \citealt{2017ApJ...837L...5A}) and PKS 1351-018 ($z$ = 3.72, \citealt{2021ApJ...919...40P}). Therefore, as an identified $\gamma$-ray emitter, B3 1239+376 ($z$ = 3.82) is proposed as the third most distant $\gamma$-ray-emitting blazar so far. Most of these extreme sources are FSRQs with soft $\gamma$-ray spectra, but NVSS J121915+365718 is suggested as a BL Lac \citep{2020ApJ...903L...8P}. B3 1239+376, along with other high-$z$ $\gamma$-ray sources, are plotted in a $\rm L_{\gamma}$-$\rm \Gamma_{\gamma}$ diagram (i.e., Figure \ref{comp}). The information listed in 4LAC-DR3 \citep{2022ApJS..263...24A}, as well as those reported in individual studies, are adopted. Considering the high apparent luminosity (up to $\rm 5\times 10^{48}$ $\rm erg~s^{-1}$ in the half-year high flux state) and relatively soft spectrum ($\rm \Gamma_{\gamma} = 2.6$), it is not surprising that our target lies with the region occupied by the high-$z$ sources. In addition to the significant variations of the flux level, the spectral hardening when brighten is indicated for some sources. However, the investigation of spectral variation is not available for B3 1239+376 due to limited statistics in the first 12-year Fermi-LAT survey. Note that several distant $\gamma$-ray sources are not included in 4LAC-DR3, alike of B3 1239+376. Because the rather faint emissions, they are detectable only in the high flux state, while the background is overwhelming for the majority of the time. Therefore, a detailed temporal analysis is crucial to catch the $\gamma$-ray signal from high-$z$ blazars.

Our jet modeling analyses suggest that B3 1239+376 is likely a low-synchrotron-peaked source (i.e., $\rm \nu_{syn} < 10^{14}$ Hz, \citealt{1995ApJ...444..567P,2010ApJ...716...30A}). In the high flux state, the synchrotron bump rises and shifts close to the infrared domain, resulting the increase of the flux densities there. On the other hand, the high energy bump moves into the detection energy range of Fermi-LAT at the same time, which leads to the emergence of the $\gamma$-ray signal. Theoretically, the injection of fresh emitting electrons in the dissipation region, as well as the enhancement of the magnetic field intensity, could account for the broadband variations. Therefore, it is not surprising that input parameters such as $\rm \gamma_{br}$ and B for the high flux state SED are larger than that of the other SED. Recent multi-wavelength campaign studying on high-$z$ blazars beyond redshift of 4 also show that injection of a hard-spectrum electron population could explain the broadband flare events \citep{2024ApJ...974...38G,2025ApJ...990..206G}. Interestingly, even in the low flux state, the X-ray emission of B3 1239+376 is luminous, which is also seen in several $\gamma$-ray quiet high-$z$ blazars \citep{2020ApJ...889..164M,2025ApJ...979L...6M}. Together with the hard X-ray spectra, blazars are believed as powerful MeV beacons in the early cosmos.

The Compton dominance (CD) of B3 1239+376 is indicated as $\sim$ 10, consistent with that found in other high-$z$ blazars \citep[e.g.,][]{2011MNRAS.411..901G,2017ApJ...837L...5A,2025ApJ...990..206G}. In particular, the CD values of B3 1428+422 and GB 1508+5714 are given as 16 and 23, respectively \citep{2024ApJ...974...38G,2025ApJ...990..206G}. CD could be an important probe on the accretion efficiency of the central SMBH, since the dense external photon fields from the accretion system could significantly boost the energy release in the $\gamma$-ray domain \citep{2017A&A...606A..44N}. In the scheme of the blazar sequence \citep{1998MNRAS.299..433F, 2011ApJ...740...98M, 2022Galax..10...35P}, more luminous blazars tend to have lower synchrotron peak frequencies and higher CD values. Majority of the hi-$z$ blazars, including B3 1239+376, fit this profile. However, note that detection of a high-$z$ $\gamma$-ray-emitting BL Lac has been reported, which possesses a relatively high-frequency peaked and a low-Compton dominated SED \citep{2020ApJ...903L...8P}. The construction of an unbiased sample at the high luminosity end of blazars is helpful to critically examine the sequence \citep{2012MNRAS.420.2899G,2021MNRAS.505.4726K}. 

Meanwhile, an important jet parameter, the bulk Lorentz factor ($\Gamma$), is worthy to be mentioned specially. In the classic orientation based AGN unified scheme, the number ratio of the jetted AGNs between the aligned and the apparent populations is suggested as $\rm 2\Gamma^{2}$. Thus, $\Gamma$ is a key parameter to determine the number density of jetted AGNs in the early cosmic time. The $\Gamma$ value for B3 1239+376 obtained here is $\sim$ 15, in agreement with the results (i. e., $\sim$ 10$-$20) provided for other similar $\gamma$-ray sources \citep[e.g.,][]{2011MNRAS.411..901G,2017ApJ...837L...5A,2020ApJ...898L..56L,2024ApJ...974...38G,2025ApJ...990..206G}.  \cite{2020ApJ...889..164M} has carried out an analysis of SEDs of a sample of high-z blazars detected with NuSTAR, though in which non-simultaneous data were adopted, also suggesting $\Gamma$ $\sim$ 10$-$15.  In addition to the jet SED modeling studies, other approaches are used to probe the jet Lorentz factor of high-$z$ sources. Searching for the potential evidence of fast variability, especially in $\gamma$ rays, can put a severe constraint on the $\Gamma$ value \citep[e.g.,][]{2018ApJ...853..159L}. Moreover, detections of the proper motion of the radio jets provide a direct estimation of the essential properties of the jet \citep[e.g.,][]{2020SciBu..65..525Z,2024MNRAS.530.4614K,2024A&A...689A..43B}. Considering the recent $\gamma$-ray activity of B3 1239+376 revealed by Fermi-LAT, quick follow-up radio interferometry observations that would reveal its jet kinematics are strongly encouraged.

In summary, based on detailed multi-wavelength investigations, B3 1239+376 ($z$ = 3.82) is identified as a $\gamma$-ray emitter, and hence it is concluded as the {\it third} most distant $\gamma$-ray-emitting blazar so far, underneath B3 1428+422 ($z$ = 4.72) and GB 1508+5714 ($z$ = 4.3). Analysis of Fermi-LAT data yields a significantly detected (7.7$\sigma$ globally) $\gamma$-ray source cospatial with B3 1239+376 in a half-year period of 2025. It is known as the unique blazar (candidate) within the location uncertainty and an association probability of 0.91 is estimated, which suggest a physical connection between the $\gamma$-ray source and B3 1239+376. Meanwhile, indications of prior $\gamma$-ray activities are also found. Although for the two epochs each alone the statistics are limited, a joint analysis returns a boosted detection significance and the B3 1239+376 remains within the $\gamma$-ray localization error radius. Moreover, in a multi-wavelength perspective, the $\gamma$-ray and infrared fluxes contemporaneously brighten. In particular, emissions in these two windows reach up to the maximal level at the same time, under a total 16-year monitoring. The temporal coincidence helps us to determine the association relationship. Interestingly, accompanying radio core flux increase until Mar. 2018 is observed. In addition to the temporal behaviors, broadband SEDs of B3 1239+376 are plotted and described by a one-zone leptonic radiation scenario. Ejections of fresh emitting electrons into the dissipation region together with an increase of the magnetic field intensity may be responsible for the variations. The soft $\gamma$-ray spectrum, luminous X-ray and $\gamma$-ray emissions, and variability across different bands found in B3 1239+376, are consistent with that found in other similar sources. To further investigate its jet properties, future multi-wavelength campaigns are urged.

\begin{acknowledgments}
This research has made use of data obtained from the High Energy Astrophysics Science Archive Research Center (HEASARC), provided by $\rm NASA^{\prime}$s Goddard Space Flight Center. This paper employs a list of Chandra datasets, obtained by the Chandra X-ray Observatory, contained in the Chandra Data Collection ~\dataset[DOI: 10.25574/cdc.567]{https://doi.org/10.25574/cdc.567}. This research makes use of data products from the Wide-field Infrared Survey Explorer, which is a joint project of the University of California, Los Angeles, and the Jet Propulsion Laboratory/California Institute of Technology, funded by the National Aeronautics and Space Administration. This research also makes use of data products from NEOWISE-R, which is a project of the Jet Propulsion Laboratory/California Institute of Technology, funded by the Planetary Science Division of the National Aeronautics and Space Administration. This publication makes use of data products from the Spectro-Photometer for the History of the Universe, Epoch of Reionization and Ices Explorer (SPHEREx), which is a joint project of the Jet Propulsion Laboratory and the California Institute of Technology, and is funded by the National Aeronautics and Space Administration. This study use data based on observations obtained with the Samuel Oschin Telescope 48-inch and the 60-inch Telescope at the Palomar Observatory as part of the Zwicky Transient Facility project. ZTF is supported by the National Science Foundation under Grant No. AST-2034437 and a collaboration including Caltech, IPAC, the Weizmann Institute for Science, the Oskar Klein Center at Stockholm University, the University of Maryland, Deutsches Elektronen-Synchrotron and Humboldt University, the TANGO Consortium of Taiwan, the University of Wisconsin at Milwaukee, Trinity College Dublin, Lawrence Livermore National Laboratories, and IN2P3, France. Operations are conducted by COO, IPAC, and UW. The National Radio Astronomy Observatory is a facility of the National Science Foundation operated under cooperative agreement by Associated Universities, Inc. CIRADA is funded by a grant from the Canada Foundation for Innovation 2017 Innovation Fund (Project 35999), as well as by the Provinces of Ontario, British Columbia Alberta, Manitoba and Quebec.

This work was supported in part by the NSFC under grants U2031120. This work was also supported in part by the Guizhou Provincial Science and Technology Projects (No. QKHFQ[2023]003 and No. QKHPTRC- ZDSYS[2023]003) and Guizhou Provincial Major Scientific and Technological Programs XKBF (2025)010 and XKBF (2025)011.
\end{acknowledgments}

%\begin{contribution}
%%This section gives authors the space to recognize author contributions. The text inside this environment is NOT counted towards the total word quanta. At a minimum, manuscripts are expected to include this text:

%All authors contributed equally to the Terra Mater collaboration.

%% But authors are expected to provide more specific details, e.g. 
%%
%%SC was responsible for writing and submitting the manuscript.
%%WWM came up with the initial research concept and edited the manuscript.
%%OTS obtained the funding and edited the manuscript.
%%EBF provided the formal analysis and validation. He also edited the manuscript.
%%GEH Supervised the undergraduates, wrote the software and administers the project github and Zenodo repositories.
%%
%% Authors can use the Contributor Role Taxonomy (CRediT) at
%% https://credit.niso.org
%% for ideas on how write a good statement tailored to their needs.

%\end{contribution}

%% To help institutions obtain information on the effectiveness of their 
%% telescopes the AAS Journals has created a group of keywords for telescope 
%% facilities.
%
%% Following the acknowledgments section, use the following syntax and the
%% \facility{} or \facilities{} macros to list the keywords of facilities used 
%% in the research for the paper.  Each keyword is check against the master 
%% list during copy editing.  Individual instruments can be provided in 
%% parentheses, after the keyword, but they are not verified.
\facilities{Fermi(LAT), CXO(ACIS), Swift, ZTF, WISE, SPHEREx, VLASS}

%% Similar to \facility{}, there is the optional \software command to allow 
%% authors a place to specify which programs were used during the creation of 
%% the manuscript. Authors should list each code and include either a
%% citation or url to the code inside ()s when available.
\software{astropy \citep{2022ApJ...935..167A}}

%% Appendix material should be preceded with a single \appendix command.
%% There should be a \section command for each appendix. Mark appendix
%% subsections with the same markup you use in the main body of the paper.
%%
%% Each Appendix (indicated with \section) will be lettered A, B, C, etc.
%% The equation counter will reset when it encounters the \appendix
%% command and will number appendix equations (A1), (A2), etc. The
%% Figure and Table counter will not reset.

%\appendix

%% For this sample we use BibTeX plus aasjournalv7.bst to generate the
%% the bibliography. The sample7.bib file was populated from ADS. To
%% get the citations to show in the compiled file do the following:
%%
%% pdflatex sample7.tex
%% bibtext sample7
%% pdflatex sample7.tex
%% pdflatex sample7.tex

\bibliography{references}{}
\bibliographystyle{aasjournal}

%% This command is needed to show the entire author+affiliation list when
%% the collaboration and author truncation commands are used.  It has to
%% go at the end of the manuscript.
%\allauthors

%% Include this line if you are using the \added, \replaced, \deleted
%% commands to see a summary list of all changes at the end of the article.
%\listofchanges

\clearpage

\begin{deluxetable}{lccccccccc}
\scriptsize
\tablenum{1} \tablewidth{0pt}
\tablecaption{Input parameters for buliding the theoretical jet SEDs in Figure \ref{sed}}
\tablehead{ \colhead{Model} &\colhead{$p_{1}$} &\colhead{$p_{2}$} &\colhead{$\gamma_{br}$} &\colhead{$K$[$\rm cm^{-3}$]} &\colhead{$B$[Gauss]} &\colhead{$\delta$} }
\startdata
High-flux state &2.0 &3.8 &387 &$\rm 4.7\times10^{4}$ &2 &15 \\[3pt]
Low-flux state &2.0 &3.8 &86 &$\rm 1.2\times10^{5}$ &1 &15  \\[3pt]
\enddata
\tablecomments{$p_{1,2}$ are the indexes of the emitting electron distribution; $\gamma_{br}$ is the electron break energy; $K$ is the normalization of the electron number density; $B$ is the magnetic field strength and $\delta$ is the Doppler boosting factor. The minimum and maximum energies of the electrons are set as 1 and 10 times of the $\gamma_{br}$, respectively.  \tiny} 
\label{tpara}
\end{deluxetable}
\end{document}